\documentstyle[prl,aps,epsfig]{revtex}

\begin{document}

\title{%
\vskip-23pt
\vskip-6pt \hfill {\rm\normalsize MPI-PhT/99-10} \\ 
\vskip-6pt \hfill {\rm\normalsize OUAST/99/9} \\ 
\vskip-6pt \hfill {\rm\normalsize March 1999} \\
Dark matter annihilation at the galactic center
}

\author{Paolo Gondolo}
\address{Max Planck Institut f\"ur Physik, F\"ohringer Ring 6,
  80805 Munich, Germany\\{\rm Email: \tt gondolo@mppmu.mpg.de}}

\author{Joseph Silk}
\address{Astrophysics, University of Oxford, Keble Road,
Oxford, OX1 3RH, U.K.\\
{\rm and}\\Department of Astronomy and Physics, 
University of California, Berkeley, CA 94720\\
{\rm Email: \tt silk@astro.ox.ac.uk}}

\twocolumn[\hsize\textwidth\columnwidth\hsize\csname@twocolumnfalse\endcsname
\maketitle
\begin{abstract}
  If cold dark matter is present at the galactic center, as in current models
  of the dark halo, it is accreted by the central black hole into a dense
  spike. Particle dark matter then annihilates strongly inside the spike,
  making it a compact source of photons, electrons, positrons, protons,
  antiprotons, and neutrinos. The spike luminosity depends on the density
  profile of the inner halo: halos with finite cores have unnoticeable spikes,
  while halos with inner cusps may have spikes so bright that the absence of a
  detected neutrino signal from the galactic center already places interesting
  upper limits on the density slope of the inner halo.  Future neutrino
  telescopes observing the galactic center could probe the inner structure of
  the dark halo, or indirectly find the nature of dark matter.
\end{abstract}
\vskip 2pc]\narrowtext

The evidence is mounting for a massive black hole at the galactic center. Ghez
et al.~\cite{ghe98} have confirmed and sharpened the Keplerian behavior of the
star velocity dispersion in the inner 0.1 pc of the galaxy found by Eckart and
Genzel~\cite{eck9697}. These groups estimate the mass of the black hole to be $
M = 2.6 \pm 0.2 \times 10^6 \, M_{\odot}$.

If cold dark matter is present at the galactic center, as in current models of
the dark halo, it is redistributed by the black hole into a cusp. We call it
the central `spike,' to avoid confusion with the inner halo cusp favored by
present N-body simulations of galaxy formation~\cite{nbody}.  If cold dark
matter contains neutral elementary particles that can annihilate with each
other, like the supersymmetric neutralino, the annihilation rate in the spike
is strongly increased as it depends on the square of the matter density. The
steep spike profile, with index $\ge 3/2$, then implies that most of the
annihilations take place at the inner radius of the spike, determined either by
self-annihilation or by capture into the black hole.

Of the annihilation end-products, neutrinos escape the spike and propagate
to us undisturbed. Current limits on the neutrino emission from the galactic
center place upper limits on the slope of the inner halo. Future neutrino
telescopes may improve on these limits or bring information on the nature of
dark matter.

\section{ADIABATIC SPIKE AROUND THE CENTRAL BLACK HOLE}

We find the dark matter density profile in the region where the black hole
dominates the gravitational potential. From the data in \cite{ghe98,eck9697},
this is the region $r \lesssim R_M \simeq 0.2 \, {\rm pc}$. 
 Other masses (the
central star cluster, for example) also influence the dark matter distribution,
but since they make the gravitational potential deeper, their effect is to
increase the central dark matter density and the annihilation signals.

We work under the assumption that the growth of the black hole is
adiabatic. This assumption is supported by the collisionless behavior of
particle dark matter.
We can find the final
density after black hole formation from the final phase-space distribution
$f'(E',L')$ as
\begin{equation}
\label{rhof}
\rho'(r) = \int_{E'_{m}}^0 dE' \,
\int_{L'_{c}}^{L'_{m}} dL' \, 
\frac{ 4 \pi L' } { r^2 v_r } \, f'(E',L') \, ,
\end{equation}
with
\begin{eqnarray}
v_r & = & \left[ 2 \left( E' + \frac{GM}{r} - \frac{L'^2}{2r^2} \right)
\right]^{1/2} , \\
E'_{m} & = & - \frac{GM}{r} \, \left( 1 - \frac{ 4 R_{\rm S}}{r} \right), \\
L'_{c} & = & 2cR_{\rm S}, \\
L'_{m} & = & \left[ 2 r^2 \left( E' + \frac{GM}{r} \right) \right]^{1/2} .
\end{eqnarray}
We have neglected the contribution from unbound orbits ($E'>0$).  The lower
limit of integration $L'_{c}$, and the second factor in $E'_{m}$, are
introduced to eliminate the particles with $L < 2cR_{\rm S}$ which are captured
by the black hole. ($R_{\rm S}=2GM/c^2$ is the Schwarzschild radius.) We relate
$f'(E',L')$ to the initial phase-space distribution $f(E,L)$ through the
relations, valid under adiabatic conditions,
$f'(E',L') = f(E,L)$, $L' = L$, $I'(E',L') = I(E,L)$,
where the last two equations are the conservation of the angular momentum $L$
and of the radial action $I(E,L)$.

The density slope in the spike depends not only on the slope of the inner halo
but also on the behavior of the initial phase-space density $f(E,L)$ as $E$
approaches the potential at the center $\phi(0)$~\cite{qui95}. If $f(E,L)$
approaches a constant, as in models with finite cores, the spike slope is
$\gamma_{\rm sp} = 3/2$.  If $f(E,L)$ diverges as $[E-\phi(0)]^{-\beta}$, as in
models with an inner cusp, the spike slope is $\gamma_{\rm sp}>3/2$. Models
with finite cores include~\cite{models}: the non-singular and the modified
isothermal sphere, the H\'enon isochrone model, the Plummer model, the King
models, the modified Hubble profile, the Evans power-law models with $R_c\ne0$,
and the Zhao $(\alpha,\beta,\gamma)$ models, $\rho \sim r^{-\gamma}
(1+r^{1/\alpha})^{-\beta\alpha}$, with $\gamma=0$ and $1/(2\alpha) = {\rm
  integer}$. Models with an inner cusp include~\cite{models}: the models of
Jaffe, Hernquist, Navarro-Frenk-White, the $\gamma/\eta$ models of Dehnen and
Tremaine et al., and the other Zhao models.

As an example of models with finite cores, we consider 
the isothermal sphere. It has $f(E,L) = \rho_0 
(2\pi\sigma_v^2)^{-3/2} \allowbreak \,
\exp(-E/\sigma_v^2)$. Close to the black hole, we have $E \ll \sigma_v^2$,
and $f(E,L) \simeq \rho_0 (2\pi\sigma_v^2)^{-3/2}$, a constant.  Then from
eq.~(\ref{rhof}) we easily find
\begin{equation}
\label{rhof1}
\rho'_{\rm iso}(r) = 
\frac{4\rho_0}{3\sqrt{\pi}} \, \left( \frac{GM}{ r\sigma_v^2} 
\right)^{3/2} \,
\left( 1 - \frac{4 R_{\rm S}}{r} \right)^{3/2} ,
\end{equation}
valid for $r \ll R_M \simeq 0.2 $ pc.
 The last factor comes from the capture of particles by the black hole:
the density vanishes for $r<4R_{\rm S}$. 

As an example of models with an inner cusp, we consider 
a single power law density profile, $\rho(r) = \rho_0 (r/r_0)^{-\gamma}$,
with $0 < \gamma < 2$. Its
 phase-space distribution function is
\begin{equation}
f(E,L) = \frac{\rho_0}{(2\pi\phi_0)^{3/2}} \,
\frac{\Gamma(\beta)}{\Gamma(\beta-\case{3}{2})} \,
\frac{\phi_0^\beta}{E^{\beta}} \, ,
\end{equation}
with $\beta = (6-\gamma)/[2(2-\gamma)] $ and $\phi_0 = 4 \pi G r_0^2
\rho_0/[(3-\gamma)(2-\gamma)]$. To find $f'(E',L')$, we need to solve $
I'(E',L') = I(E,L) $ for $E$ as a function of $E'$. In the field of
a point-like mass, we have 
$ 
I'(E',L') = 2 \pi \left[ - L' + GM/\sqrt{-2E'} \right] 
$.
In the field of the power law profile, whose potential is proportional to
$r^{2-\gamma}$, the action integral cannot be performed exactly. We have
found an approximation good to better than 8\% over all of phase space for
$0 < \gamma < 2$: 
\begin{equation}
I(E,L) = \frac{2\pi}{b} \, \left[ - \frac{L}{\lambda} + 
\sqrt{2 r_0^2 \phi_0} \left( \frac{E}{\phi_0} \right) ^ 
{ \frac{4-\gamma}{2(2-\gamma)} } \right],
\end{equation}
where $\lambda = [2/(4-\gamma)]^{1/(2-\gamma)} \,
[(2-\gamma)/(4-\gamma)]^{1/2}$ and $b = \pi(2-\gamma)/{\rm
  B}(\frac{1}{2-\gamma},\frac{3}{2}) $.  Expressing $E$ as a function of $E'$
and integrating eq.~(\ref{rhof}), we obtain
\begin{equation}
\label{rhoprime}
\rho'(r) = \rho_R \, g_{\gamma}(r) \,
\left( \frac{R_{\rm sp}}{r}\right)^{\gamma_{\rm sp}} ,
\end{equation}
with 
$ \rho_R = \rho_0 \left( {R_{\rm sp}}/{r_0} \right)^{-\gamma} $,
$ \gamma_{\rm sp} = (9-2\gamma)/(4-\gamma) $, and
$ R_{\rm sp} = \alpha_{\gamma} r_0 \left( M/\rho_0 r_0^3
\right)^{1/(3-\gamma)} $.
For $0\le \gamma \le 2$, the density slope in the spike, $\gamma_{\rm sp}$,
varies only between 2.25 and 2.5. 
\begin{figure}[t]
\label{figprofile}
\epsfig{file=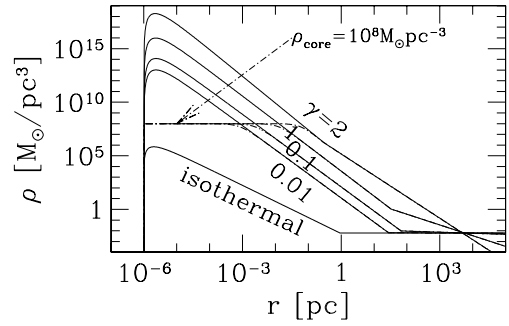,width=0.4\textwidth}
\caption{
  Examples of spike density profiles.}
\end{figure}
 While the exponent $\gamma_{\rm sp}$ can
also be obtained by scaling arguments~\cite{qui95}, the normalization
$\alpha_{\gamma}$ and the factor $g_{\gamma}(r)$ accounting for capture must be
obtained numerically. We find that $g_{\gamma}(r) \simeq ( 1 -4 R_{\rm S}/r)^3$
over our range of $\gamma$, and that $\alpha_{\gamma} \simeq 0.293
\gamma^{4/9}$ for $\gamma \ll 1$, and is $\alpha_{\gamma}=$ 0.00733, 0.120,
0.140, 0.142, 0.135, 0.122, 0.103, 0.0818, 0.0177 at
$\gamma=0.05,0.2,0.4,\ldots,1.4,2$.  The density falls rapidly to zero at $r
\lesssim 9.55 R_{\rm S}$, vanishing for $r<4R_{\rm S}$.

Annihilations in the inner regions of the spike set a maximal dark matter
density $\rho_{\rm core} = m/\sigma v t_{\rm bh}$, where $t_{\rm bh}$ is the
age of the black hole, conservatively $10^{10}$ yr, $m$ is the mass of the dark
matter particle, and $\sigma v$ is its annihilation cross section times
relative velocity (notice that for non-relativistic particles $\sigma v$ is
independent of $v$). Using $\partial \rho/\partial t = - \sigma v \rho^2/m$,
the final spike profile is
\begin{equation}
\rho_{\rm sp}(r) = \frac{ \rho'(r) \rho_{\rm core} } { \rho'(r) + \rho_{\rm
    core} } ,
\end{equation}
which has a core of radius $R_{\rm core} = R_{\rm sp} \left( \rho_R/\rho_{\rm
    core} \right) ^ {1/\gamma_{\rm sp}} $. In the particle models we consider,
not more than the initial amount of dark matter within 300 pc is annihilated.

Examples of spike density profiles are shown in Fig.~1.

To conclude this section, we derive a conservative estimate of the dark matter
density near the galactic center.  Assume that the halo density is constant on
concentric ellipsoids.  Then the halo contribution $v_{h}(r)$ to the rotation
speed at distance $r$ fixes the halo mass within $r$. Assume further that the
halo density decreases monotonically with distance. Since at large radii it
decreases at least as fast as $r^{-2}$, and at small radii only as
$r^{-\gamma}$ with $\gamma < 2$, the density profile becomes steeper with
distance. So to continue the $r^{-\gamma}$ dependence to all radii keeping the
same halo mass interior to $r$ as given by the rotation speed, we must decrease
the density normalization. In this way we obtain an underestimate of the
density near the center. Letting $\rho_D = \rho_0 (D/r_0)^{-\gamma} $, we have
\begin{equation}
\frac{ \rho_D }{ 1 - \gamma/3} \sim 
\frac{ 3 v^2_h(D) }{ 4 \pi G D^2 } 
\simeq 0.0062 \, \frac{ M_{\odot} }{ {\rm pc}^{3} }
\simeq 0.24 \, \frac{ {\rm GeV} }{ {\rm cm}^{3} },
\end{equation}
where we have taken $v_{h}(D) = 90 {\rm \,km\,s^{-1}}$ at the Sun distance $D
= 8.5$ kpc, as obtained after subtracting a (somewhat overestimated) luminous
matter contribution of 180 km s$^{-1}$ to the circular speed of $220 \pm 20
{\rm \,km\,s^{-1}}$~\cite{deh98}.

\section{SIGNALS FROM NEUTRALINO ANNIHILATIONS}

Our analysis applies in general to a self-annihilating dark matter particle. To
make it concrete, we examine a case of supersymmetric dark matter, the lightest
neutralino.  The minimal supersymmetric standard model privides a well-defined
calculational framework, but contains at least 106 yet-unmeasured
parameters~\cite{dim95}. Most of them control details of the squark and slepton
sectors, and can safely be disregarded in dark matter studies. So we restrict
the number of parameters to 7, following Bergstr\"om and Gondolo~\cite{ber96}.
Out of the database of points in parameter space built in
refs.~\cite{ber96,eds97,ber98}, we use the 35121 points in which the neutralino
is a good cold dark matter candidate, in the sense that its relic density
satisfies $0.025 < \Omega_\chi h^2 < 1 $.  The upper limit comes from the age
of the Universe, the lower one from requiring that neutralinos are a major
fraction of galactic dark halos.

Gravitational interactions bring the cold neutralinos into our galactic halo
and into the central spike, where neutralino pairs can annihilate and produce
photons, electrons, positrons, protons, antiprotons, and neutrinos. While most
products are subject to absorption and/or diffusion, the neutrinos escape
the spike and propagate to us undisturbed. We focus on the neutrinos and
postpone the study of other signals.

The expected neutrino flux from neutralino annihilations in the direction
of the galactic center can be divided into two components: emission from the
halo along the line of sight and emission from the central spike,
\begin{equation}
\Phi_{\nu}^{\rm neutralinos} =
\Phi_{\nu}^{\rm halo} + \Phi_{\nu}^{\rm spike} .
\end{equation}

The halo flux from neutralino annihilations between us and the galactic center
can be estimated assuming that a single power law profile
$\rho(r) = \rho_D (r/D)^{-\gamma} $ extends out to the Sun position $r=D$. 
The integrated neutrino flux within an angle $\Theta$ of the
galactic center is
\begin{equation}
\label{phihalo}
\Phi_{\nu}^{\rm halo} = 
\frac{ \rho_D^2 Y_{\nu} \sigma v D \Omega(\Theta) }{ m^2 } ,
\end{equation}
where 
$m$ is the neutralino mass, $Y_{\nu}$ is the number of neutrinos
produced per annihilation, either differential or integrated in energy, 
$\sigma v $ is the
neutralino--neutralino annihilation cross section times relative velocity,
and
\begin{equation}
\Omega(\Theta) =  
\frac{\Theta^2 }{2(1-2\gamma)} - \frac{2^{\gamma-1/2}
\Theta^{3-2\gamma}}{(3-2\gamma)(1-2\gamma)} -
\frac{ \Theta_{\rm min}^{3-2\gamma}}{3-2\gamma} .
\end{equation}
Here $\Theta$ is in radians and $\Theta_{\rm min} = \max \bigl[ 10R_{\rm S}/D ,
$ $ (2\gamma/3)^{1/(3-2\gamma)} (\rho_D/\rho_{\rm
  core})^{1/\gamma} \bigr]$.

We evaluate the neutrino yield $Y_\nu$ and the neutralino annihilation cross
sections $\sigma$ using the DarkSUSY package~\cite{gon99}, which incorporates
Pythia simulations of the $\nu$ continuum~\cite{ber99}
and the annihilation cross sections in~\cite{eds97,annih}.
\begin{figure}[t]
\label{figenh}
\epsfig{file=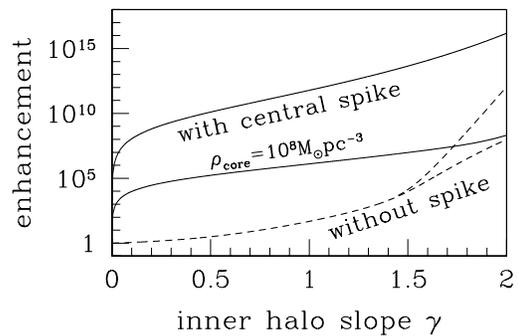,width=0.4\textwidth}
\caption{
  Enhancement of annihilation signals from the
  galactic center.}
\end{figure}
\begin{figure}[tbp]
\label{figphimu}
\epsfig{file=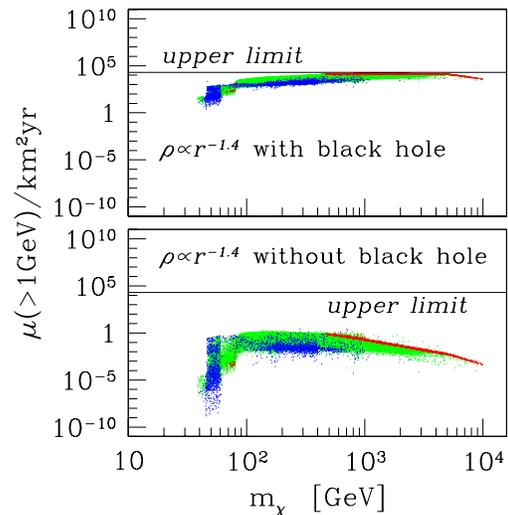,width=0.4\textwidth}
\caption{For a Moore et al.\ halo
  profile, flux of neutrino-induced muons in a neutrino telescope from
  neutralino dark matter annihilations in the direction of the galactic center,
  with (upper panel) and without (lower panel) the central spike. The
  horizontal line is the current upper limit.  }
\end{figure}

To the halo flux we need to add the contribution from the spike around the
black hole at the galactic center. 
For an isothermal distribution 
we find
\begin{equation}
\label{phispike1}
\Phi_{\nu}^{\rm spike} =
\frac{ \rho_D^2 Y_{\nu} \sigma v D} {m^2} \,
\left( \frac{ R_M }{ D } \right)^{3} \,
\, 
\ln \left( \frac{ R_M }{ 25 R_{\rm S} } \right) ,
\end{equation}
the factor of 25 serving to match the exact integration.  This flux is a factor
of $\sim 10^{-9}$ smaller than the flux from dark matter annihilations between
us and the galactic center, and so the addition of the spike does not modify
the signal.  The same conclusion is reached in general for halo models with
finite cores.

A strong enhancement results instead 
for power law profiles. We find
\begin{equation}
\label{phispike2}
\Phi_{\nu}^{\rm spike} = 
\frac{ \rho_D^2 Y_{\nu} \sigma v D} {m^2} \,
\left( \frac{ R_{\rm sp} }{ D } \right)^{3-2\gamma} \,
\left( \frac{ R_{\rm sp} }{ R_{\rm in} } \right)^{2\gamma_{\rm sp}-3} ,
\end{equation}
where $R_{\rm sp}$ is given after eq.~(\ref{rhoprime}) with $\rho_0$
replaced by $\rho_D$ and $r_0$ by $D$.
We fix $R_{\rm in}$ so as to match the
integration of the numerically-calculated density profile including capture and
annihilation: we
find that $R_{\rm in} = 1.5 \left[ (20 R_{\rm S})^2+R_{\rm core}^2
\right]^{1/2} $ gives a good approximation to the flux
(within 6\% for our values of $\gamma$).

Contrary to the case of finite cores, for cusped halos there is a huge increase
in flux from the galactic center when the spike is included, typically 5 orders
of magnitude or more, unless the inner halo slope $\gamma$ is very small (see
Fig.~2, where $\Theta=1.\!\!^\circ 5$).  The enhancement is notable, for
example, for the profile of Moore et al.\ which has $\gamma=1.4$ (see
Fig.~3): including the spike around the black hole dramatically changes 
the prospects of observing a neutrino flux.

The neutrino flux from the spike increases with the inner halo slope $\gamma$.
Imposing that it does not exceed the observed upper limit of $\sim 2 \times
10^4$ muons($>\!1$GeV) km$^{-2}$ yr$^{-1}$ \cite{imb-kgf-kamiokande} leads to
an upper bound on $\gamma$. There is a separate upper bound for each model.
They are plotted in Fig.~4a. (Plotted values of $\gamma_{\rm max}>2$ are
unphysical extrapolations but are shown for completeness.) Present bounds are
of the order of $\gamma_{\rm max} \sim 1.5$.

Future neutrino telescopes situated in the Northern hemisphere may improve on
this bound or find a signal. For example, with a muon energy threshold of 25
GeV, the neutrino flux from the spike after imposing the current constraints
could still be over 3 orders of magnitude above the atmospheric background
(Fig.~5), allowing to probe $\gamma$ as low as 0.05 (Fig.~4b).
\begin{figure}[tbp]
\label{figgmaxnu}
\epsfig{file=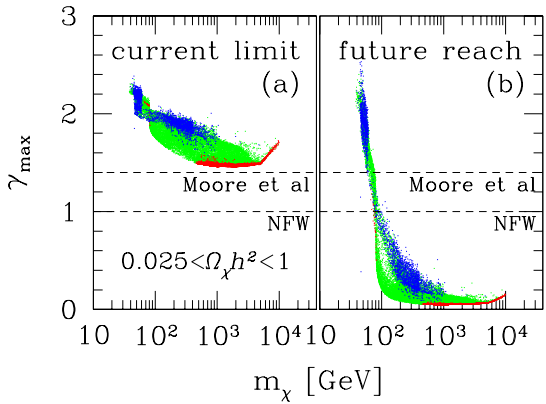,width=0.4\textwidth}
\caption{Maximum inner slope
  $\gamma$ of the dark matter halo compatible with the upper limit on the
  neutrino emission from the galactic center. (a) Current limit at 1 GeV;
  (b) future reach at 25 GeV.}
\end{figure}

In conclusion, we have shown that if the galactic dark halo is cusped, as
favored in recent N-body simulations of galaxy formation, a bright dark matter
spike would form around the black hole at the galactic center. A search of a 
neutrino signal from the spike could either set upper bounds on the
density slope of the inner halo or clarify the nature of dark matter. 

This research has been supported in part by grants from NASA and DOE.

\begin{figure}[tbp]
\label{figphimumax}
\epsfig{file=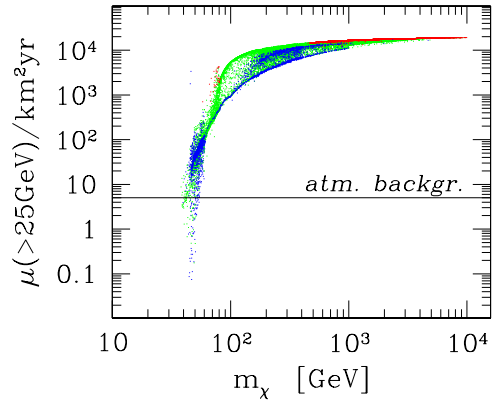,width=0.4\textwidth}
\caption{Maximal flux of neutrino-induced muons in a neutrino telescope from
  neutralino annihilations at the galactic center, after imposing the current
  constraints on the neutrino emission.}
\end{figure}

\end{document}